\documentclass[12pt]{article}% michael
\usepackage{times}% manny
\usepackage[dvips]{graphicx,color}% manny
\newcommand{\ignore}[1]{}

\newcommand{\mComment}[1]{}
\newcommand{\nComment}[1]{}
\ignore{
\renewcommand{\mComment}[1]{\textcolor{blue}{Manny: #1}} %manny
\renewcommand{\nComment}[1]{\textcolor{magenta}{Michael: #1}} %manny
}

\newcommand{\ket}[1]{|{#1}\rangle}

\newcommand{\cA}{{\cal A}}

\newcommand{\cH}{{\cal H}}

\newcommand{\crdl}{\|}

\begin{document}

\textsc{\textbf{Quantum computation, theory of}}\footnote{Article by
  E.~H.~Knill and M.~A.~Nielsen accepted for Supplement III,
  Encyclopaedia of Mathematics (publication expected Summer 2001).
  See also ``http://www.wkap.nl/series.htm/ENM''.  E.~H.~Knill is with
  the Los Alamos National Laboratory, MS B265, Los Alamos NM 87545,
  USA, and M.~A.~Nielsen is with the Center for Quantum Computer
  Technology, Department of Physics, University of Queensland 4072,
  Australia.  } - The study of the model of computation in which the
state space consists of linear superpositions of classical
configurations and the computational steps consist of applying local
unitary operators and measurements as permitted by quantum mechanics.

Quantum computation emerged in the 1980's when P.~Benioff and
R.~Feynman realized that the apparent exponential complexity in
simulating quantum physics could be overcome by using a sufficiently
well controlled quantum mechanical system to perform a simulation.
Quantum Turing machines were introduced by D.~Deutsch in 1985.
Initial work focused on how quantum mechanics could be used to
implement classical computation (computation in the sense of A.~Church
and A.~Turing), and on analyzing whether the quantum Turing machine
model provided a universal model of computation.  In the early 1990's,
D.~Deutsch and R.~Jozsa found an oracle problem that could be solved
faster on an error-free quantum computer than on any deterministic
classical computer.  E.~Bernstein and U.~Vazirani then formalized the
notion of quantum complexity from a theoretical computer
science point of view, and showed that with respect to oracles which
reversibly compute classical functions, quantum computers are
super-polynomially more efficient than classical computers.  The gap
was soon improved to an exponential one. This work culminated in
P.~Shor's discovery of an efficient (that is, consuming only
polynomial resources) algorithm for factoring large numbers and for
computing discrete logarithms. It implied that widely used public key
cryptographic systems would be insecure if quantum computers were
available.  Subsequently, L.~Grover found an algorithm which permitted
a square-root speed-up of unstructured search. Finding new algorithmic
improvements achievable with quantum computers which are not reducible
to Shor's or Grover's algorithm is currently (2000) an active research
area.  Also of great current interest is understanding how the problem
of simulating quantum systems, known to be tractable on a quantum
computer, relates to the problems conventionally studied within
classical computational complexity theory.  Comprehensive
introductions to quantum computation and the known quantum algorithms
may be found in \cite{Nielsen00a,Gruska99a}.

The algorithmic work described above firmly established the field of
quantum computation in computer science. However, it was initially
unclear whether quantum computation was a physically realizable model.
Particularly worrisome was the fact that in nature, quantum effects
are rarely observable, and in fact, physical noise processes tend to
rapidly remove the necessary phase relationships. To solve the problem
of quantum noise, P.~Shor and A.~Steane introduced quantum
error-correcting codes. This idea was expanded and applied by
several research groups to prove that under physically reasonable
assumptions, fault tolerant quantum computation is possible. Among the
assumptions are the requirements that quantum noise is sufficiently
weak (below some constant threshold error per quantum bit and
operation) and that the basic operations can be performed in parallel.
As a result there are now many intense experimental efforts devoted
toward realizing quantum computation, in a wide and increasing variety
of physical systems.  Progress to date (2000) has been modest, with
existing systems limited to just a few qubits, and on the order of one
hundred operations~\cite{braunstein:qc2000b}.

Models of quantum computation largely parallel and generalize the
classical models of computation. In particular, for formal studies of
complexity, many researchers use various versions of quantum Turing
machines, while quantum random access machines or quantum networks
(also known as quantum circuits) are preferred for describing and
investigating specific algorithms.  To obtain a quantum version of a
classical model of deterministic computation, one begins with the
classical model's state space. The classical state space usually
consists of an enumerable set of configurations $\psi_i$, with index
$i$ often constructed from strings of symbols.  The quantum model
associates to each $\psi_i$ a member of a standard orthonormal basis
$\ket{i}$ (called \emph{classical states}) of a Hilbert space $\cH$.
The states of the quantum model are given by ``superpositions'' of
these basis states, which are unit vectors in $\cH$.  The classical
model's initial state $\psi_0$ becomes the quantum model's initial
state $\ket{0}$, and the classical model's transition function is
replaced by a unitary operator $U$ acting on $\cH$. $U$ has to satisfy
certain locality restrictions that imply, for example, that
$U\ket{i}$ must be a superposition of classical states that are
accessible by an allowed classical transition function in one step
from $\psi_i$.  The computation's answer can be obtained by
\emph{measuring} the state after each step.  In the simplest case, the
classical computation's answer is determined by whether the
configuration is an ``accepting'' one.  Accepting configurations form
a set $\cA$ which may be associated with the closed subspace of $\cH$
spanned by the corresponding classical states. Let $P$ be the
projection operator onto this subspace. If the state of the quantum
model is $\ket{\phi}$, measurement has two possible outcomes. Either
the new state is $P\ket{\phi}/\crdl P\ket{\phi}\crdl$ with probability
$p=\crdl P\ket{\phi}\crdl^2$, in which case the computation
``accepts'', or the state is $(1-P)\ket{\phi}/\crdl
(1-P)\ket{\phi}\crdl$ with probability $1-p$, in which case the
computation continues.  The possible measurement outcomes can be
expanded by adding a set of ``rejecting'' states.  In the early days
of quantum computation there were lively discussions of how quantum
Turing machines should halt, implying different rules about when
measurements are applied during a computation.

The method outlined above for obtaining a quantum model of computation
from a classical model yields a generalization of the restriction of
the classical model to reversible transition functions. This implies
that quantum complexity classes do not necessarily enlarge the
classical analogues, particularly for the low-lying classes or when
restricted models of computation (for example, finite state automata)
are involved.  To obtain a generalization of the usual model of
computation it suffices to extend the set of transition operators with
suitable irreversible ones. One way to do that is to allow transition
operators which are the composition of a measurement (satisfying an
appropriate locality constraint) followed by unitary operators
depending on the measurement outcome.  A different approach which
works well for random access machines (RAM) is to enhance the RAM by
giving it access to an unbounded number of \emph{quantum bits} which
can be controlled by applying \emph{quantum gates} (cf.
\textbf{quantum information processing}). This is in effect how
existing quantum algorithms are described and analyzed.

As in classical complexity studies, resources considered for quantum
complexity include time and space. In the context of irreversible
processes, an additional resource that may be considered is entropy
generated by irreversible operations.  When analyzing algorithms based
on quantum RAMs, it is also useful to separately account for classical
and quantum resources.  It is important to realize that if the complex
coefficients of the unitary transition operators are rational (or in
general, \emph{computable} complex numbers), then there is no
difference between classical and quantum computability. Thus the
functions computable by quantum Turing machines are the same as those
computable by classical Turing machines.

An important issue in studying quantum models of computation is how to
define the  computation's ``answer'' given that the output is
intrinsically probabilistic. How this is defined can affect complexity
classes. Guidance comes from studies of probabilistic (or randomized)
computation, where the same issues arise. Since quantum computation
with irreversibility can also be viewed as a generalization of
probabilistic computation, most comparisons of the complexity of
algorithms use bounds on the efficiency of probabilistic algorithms.

The best known quantum complexity class is the class of bounded error
quantum polynomial time computable languages ($\mathbf{BQP}$).  This
is the class of languages decided in polynomial time with probability
$>2/3$ (acceptance) and $<1/3$ (rejection) by a quantum Turing
machine.  Based on the oracle computing studies, the quantum
factoring algorithm, and the difficulty of classically simulating
quantum physics, it is conjectured that $\mathbf{BQP}$ strictly
contains $\mathbf{BPP}$ (the class of bounded error polynomial time
computable languages for the model of probabilistic classical
computation).  $\mathbf{BQP}$ is contained in
$\mathbf{P}^{\raisebox{.5pt}{\#}\mathbf{P}}$ (the class of languages
decidable in polynomial time on a classical Turing machine given
access to an oracle for computing the permanent of $0$-$1$
matrices---this class is contained in the class $\mathbf{PSPACE}$ of
languages computable using polynomial working space).  Thus, a proof
of the important conjecture that $\mathbf{BQP}$ is strictly larger
than $\mathbf{BPP}$ will imply the long-sought result in classical
computational complexity that $\mathbf{BPP} \neq \mathbf{PSPACE}$.

The relationship of $\mathbf{BQP}$ to $\mathbf{NP}$ (the class of
nondeterministic polynomial time languages) is not known, though it is
conjectured that $\mathbf{NP}\not<\mathbf{BQP}$.  If this is not the
case, it would have immense practical significance, as many
combinatorial optimization problems are in $\mathbf{NP}$.  One reason
for thinking that $\mathbf{NP}\not<\mathbf{BQP}$ is the fact that
Grover's algorithm provides the optimal speedup for unstructured
quantum search, 
and it is widely believed that the
reason for the difficulty of solving $\mathbf{NP}$-complete problems
is that it is essentially equivalent to searching an unstructured
search space.  A generalization of unstructured search involves
determining properties of (quantum) oracles by means of queries.  In
classical computation, an oracle is a function $f$ with values in
$\{0,1\}$. The corresponding quantum oracle applies the unitary
operator $\hat f$ defined on basis states by $\hat
f\ket{x,0,w}\rightarrow\ket{x,f(x),w}$ and $\hat
f\ket{x,1,w}\rightarrow\ket{x,1-f(x),w}$.  To \emph{query} the oracle,
one applies $\hat f$ to the current state.  Grover's algorithm can be
cast in terms of an oracle problem.  The observation that this
algorithm is optimal has been extended by using the method of
polynomials~\cite{beals:qc1998a} to show that when no promise is made on
the behavior of the oracle, quantum computers are at most polynomially
more efficient than classical computers.

An area where there are provable exponential gaps between the
efficiency of quantum and classical computation occurs when
communication resources are taken into consideration.  This area is
known as quantum communication complexity (introduced by A.~Yao in
1993) and considers problems where two parties with quantum computers
and a quantum channel between them (cf. \textbf{quantum information
processing}) jointly compute a function of their respective inputs and
wish to minimize the number of quantum bits communicated.  The
exponential gaps between quantum and classical communication
complexity are so far confined to problems where the inputs to the
function computed are constrained by a ``promise''~\cite{Raz99a}.
The best known gap without a promise is a quadratic separation between
classical and quantum protocols with bounded probability of
error~\cite{buhrman:qc1998a}.  Several research groups have developed
techniques for proving lower bounds on quantum communication
complexity, mostly variations of the log-rank lower bound also used in
classical communication complexity.  These results show that for some
problems (for example, computing the inner product modulo two of bit
strings known to the respective parties) there is little advantage to
using quantum information processing.

%\bibliographystyle{plain}
%\bibliography{journalDefs,qc,qc_nielsen.bib}

\begin{center}\mbox{}\hspace*{\fill}
\begin{tabular}{l}
E. H. Knill\\
M. A. Nielsen
\end{tabular}
\end{center}


\end{document}